\documentclass[twocolumn,10pt]{asme2e}
\usepackage{graphicx}

\confshortname{FEDSM2018}
\conffullname{the ASME 2018 5th Joint US-European Fluids Engineering Summer Conference}

\confdate{July 15-20}
\confyear{2018}
\confcity{Montreal, Quebec}
\confcountry{Canada}

\papernum{FEDSM2018-83359}

\title{ROUTE TO CHAOS IN THE FLUIDIC PINBALL}

\author{Nan Deng
    \affiliation{
LIMSI \\
Universit\'e Paris-Sud / Paris-Saclay  \\
 F-91405 Orsay, France  \\
 nan.deng@u-psud.fr}	
}

\author{Luc R. Pastur\thanks{Address all correspondence to this author.} \\
    \affiliation{ IMSIA-ENSTA ParisTech \\
F-91762 Palaiseau, France \\ \\
LIMSI \\
Universit\'e Paris-Sud / Paris-Saclay \\
 F-91405 Orsay, France  \\ 
 luc.pastur@limsi.fr
    }
}
\author{Marek Morzy\'nski    
    \affiliation{Institute of Combustion Engines and Basics of Machine Design, \\ Pozna\'n University of Technology \\
PL 60-965 Pozna\'n, Poland\\
	Marek.Morzynski@put.poznan.pl \\
    }
}
\author{Bernd R. Noack\\
    \affiliation{LIMSI \\
Universit\'e Paris-Sud / Paris-Saclay  \\
     F-91405 Orsay, France \\  \\
Institute for Turbulence-Noise-Vibration Interaction and Control \\
Harbin Institute of Technology
Shenzhen, People's Republic of China \\  \\
Institut f\"ur Str\"omungsmechanik und Technische Akustik (ISTA) \\
Technische Universit\"at Berlin \\
D-10623 Berlin, Germany\\
	noack@limsi.fr
    }
}

\begin{document}

\maketitle    

\begin{abstract}
%
%
{\it The fluidic pinball has been recently proposed as an attractive and effective flow configuration for exploring machine learning fluid flow control. In this contribution, we focus on the route to chaos in this system without actuation, as the Reynolds number is smoothly increased. The route was found to be of the Newhouse-Ruelle-Takens kind, with a secondary pitchfork bifurcation that breaks the symmetry of the mean flow field on the route to quasi-periodicity. 
}

\end{abstract}

\begin{nomenclature}
\entry{$x,y,z$}{Streamwise, crosswise and transverse direction, respectively.}
\entry{$\mathbf{u}$}{Velocity flow field.}
\entry{$u,v,w$}{Streamwise, crosswise and transverse components of the velocity flow field, respectively.}
\entry{$U_\infty$}{In-coming flow velocity.}
\entry{$C_F$, $C_T$, $C_B$}{Front, top and bottom lift coefficients.}
\entry{$C_L$}{Total lift coefficient.}
\entry{$R$}{Cylinder radius. $L=5R$.}
\entry{$D$}{Cylinder diameter.}
\entry{$\rho $}{Fluid density.}
\entry{$\nu$}{Kinematic viscosity.}
\entry{$Re_D$}{Reynolds number based on $U_\infty$ and $D$.}
\entry{$t_{\mathrm{conv}}$}{Convective time expressed in units of $D/U_\infty$.}
\entry{$f$}{Frequency of the dominant peak in power spectral densities.}
\entry{$\tau$}{Time delay for phase portrait representation, chosen as the fourth of the vortex shedding dominant period $1/f$.}
\end{nomenclature}

\section*{INTRODUCTION}

Machine learning control (MLC) has been recently successfully applied to closed-loop turbulence control experiments for mixing enhancement \cite{parezanovic2016}, reduction of circulation zones \cite{gautier2015}, separation mitigation of turbulent boundary layers \cite{hu2008a,hu2008b}, force control of a car model \cite{li2016} and strongly nonlinear dynamical systems featuring aspects of turbulence control \cite{brunton2015,duriez2017}. In all cases, a simple genetic programming algorithm has learned the optimal control for the given cost function and out-performed existing open- and closed-loop approaches after few hundreds to few thousands test runs. Yet, there are numerous opportunities to reduce the learning time by avoiding the testing of similar control laws and to improve the performance measure by generalizing the considered control laws. In addition, running thousands of tests before converging to the optimal control law can be out-of-reach when dealing with heavy numerical simulations of the Navier-Stokes equations.  

In order to further improve MLC strategies, it is, therefore, of the utmost importance to handle numerical simulations of the Navier-Stokes equations in flow configurations that are geometrically simple enough to allow testing of control laws within minutes on a Laptop, while being physically complex enough to host a range of interacting frequencies. With that aim in mind, Noack \& Morzynski \cite{tutorial} proposed, as an attractive flow configuration, the uniform flow around 3 cylinders, which can be rotated around their axis (3 control inputs), with multiple downstream velocity sensors as multiple outputs. As a standard objective, the control goal could be to stabilize the wake or reduce the drag.
This configuration, proposed as a new benchmark for multiple inputs-multiple outputs (MIMO) nonlinear flow control, was named as the \textit{fluidic pinball}, as the rotation speeds allow to change the paths of the incoming fluid particles like flippers manipulate the ball of a real pinball.
Despite its geometric simplicity, this configuration exhibits a large range of flow behaviors, from steady state to chaotic dynamics, as detailed in the forthcoming part of the paper.

%

\section*{THE FLUIDIC PINBALL}

\begin{figure}
 \includegraphics[width=\linewidth]{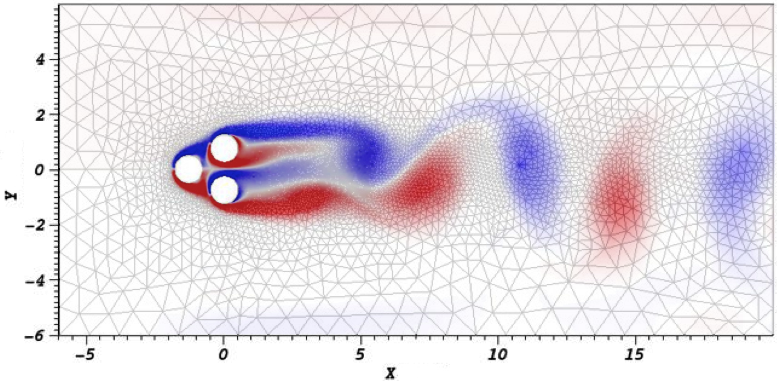}
 \caption{CONFIGURATION OF THE FLUIDIC PINBALL: THE THREE CYLINDERS ARE IN WHITE, THE FLOW IS COMING FROM THE LEFT. THE COLORMAP ENCODES AN INSTANTANEOUS VORTICITY FIELD. }
 \label{fig:config}
\end{figure}

The \textit{fluidic pinball} is made of three equal circular cylinders of radius $R$ that are placed in parallel in a viscous incompressible uniform flow with speed $U_\infty $. The centers of the cylinders form an equilateral triangle with side-length $3R$, symmetrically positioned with respect to the flow (see Fig.~\ref{fig:config}). The leftmost triangle vertex points upstream, while rightmost side is orthogonal to the on-coming flow. Thus, the transverse extend of the three cylinder configuration is given by $L = 5R$. This flow is described in a Cartesian coordinate system, where the $x$-axis points in the direction of the  flow, the $z$-axis is aligned with the cylinder axes, and the $y$-axis is orthogonal to both. The origin 0 of this coordinate system coincides with the mid-point of the rightmost bottom and top cylinder. The location is described by $\mathbf{x} = (x; y; z) = x\,\mathbf{e}_x + y\,\mathbf{e}_y + z\,\mathbf{e}_z$, where $\mathbf{e}_{x;y;z}$ are unit vectors pointing in the direction of the corresponding axes. Analogously, the velocity reads $\mathbf{u} = (u; v;w) = u\,\mathbf{e}_x +v\,\mathbf{e}_y +w\,\mathbf{e}_z$. The pressure is denoted by $p$ and time by $t$. In the following, we assume a two-dimensional flow, i.e. no dependency of any flow quantity on $z$ and vanishing spanwise velocity $w\equiv 0$. The Newtonian fluid is characterized by a constant density $\rho$ and kinematic viscosity $\nu$. In the following, all quantities are assumed to be non-dimensionalized with cylinder diameter $D = 2R$, velocity $U_\infty $ and fluid density $\rho$. The corresponding Reynolds number is defined as $Re_D = U_\infty D/\nu$. The Reynolds number based on the transverse length $L = 5D$ is 2.5 times larger. The computational domain extends from $x = -6$ up to $x = 20$ in the streamwise direction, and from $y = -6$ up to $y = 6$ in the crosswise direction. With this non-dimensionalization, the cylinder axes are located at:
$$
\begin{array}{lcl}
x_F =  - 3/2 \cos 30^{\circ}, & & y_F = 0, \\
x_B = 0, & & y_B = - 3/4, \\
x_T = 0, & & y_T = + 3/4. \\
\end{array}
$$
Here, and in the following, the subscripts `F', `B' and `T' refer to the front, bottom and top cylinder. 

The dynamics of the flow is governed by the incompressible Navier-Stokes equations:
\begin{eqnarray}
 \frac{\partial  \mathbf{u}}{\partial t} + \mathbf{u}\cdot \nabla \mathbf{u} & = & -\nabla p + \frac{1}{Re_D} \Delta \mathbf{u},  \\
 \nabla \cdot \mathbf{u} & = & 0, 
 \label{eq:ns}
\end{eqnarray}
where $\nabla $ represents the Nabla operator, $\partial _t$ and $\Delta$ denote the partial derivative and the Laplace operator. 
Without forcing, the boundary conditions comprise a no slip-condition ($\mathbf{u} = 0$) on the cylinder and a free-stream condition ($\mathbf{u} =\mathbf{e}_x$) in the far field. 
The flow can be forced by rotating the cylinders. 

The flow domain is discretized on an unstructured grid with 4\,225 triangles and 8\,633 vertices. The discretization optimizes the speed of the numerical simulation while keeping the accuracy at acceptable level. The Navier-Stokes equation is numerically integrated with an implicit Finite-Element Method \cite{noack2003,noack2016}. The numerical integration is third-order accurate in space and time. For more details about the fluidic pinball configuration, the interested reader can refer to the technical report and user manual by Noack \& Morzynski \cite{tutorial}.

\section*{NATURAL FLOW REGIMES}

\begin{figure}[tb]
 \includegraphics[width=\linewidth]{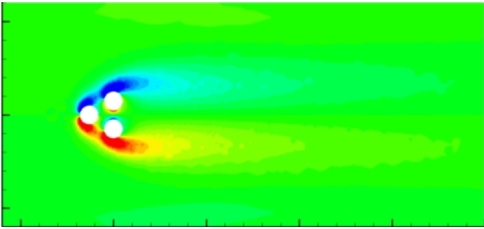}
 \caption{STEADY BASE FLOW AT $Re_D=10$. THE COLORMAP ENCODES THE VORTICITY FIELD.} 
 \label{fig:steadysolution}
\end{figure}

\begin{figure}[tb]
 \includegraphics[width=\linewidth]{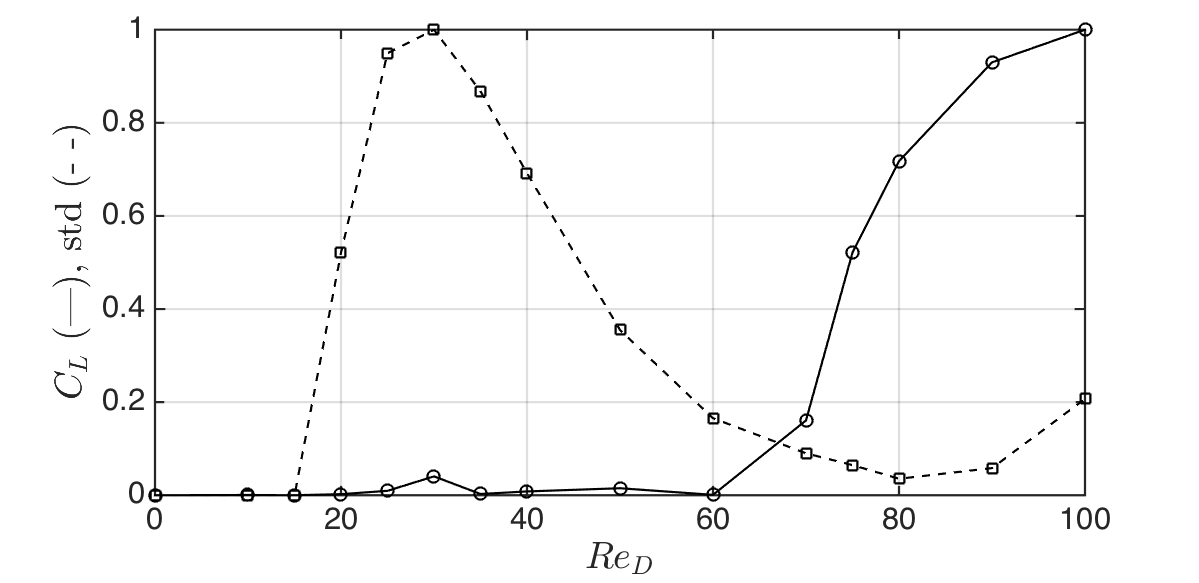}
 \caption{EVOLUTION WITH THE REYNOLDS NUMBER OF THE GLOBAL LIFT COEFFICIENT $C_L$ (SOLID LINE) AND ITS STANDARD DEVIATION (DASHED LINE), WITH MIN-MAX NORMALIZATION.}
 \label{fig:CL}
\end{figure}

\begin{figure}[tb]
\begin{tabular}{c}
 (a) \\ \includegraphics[width=\linewidth]{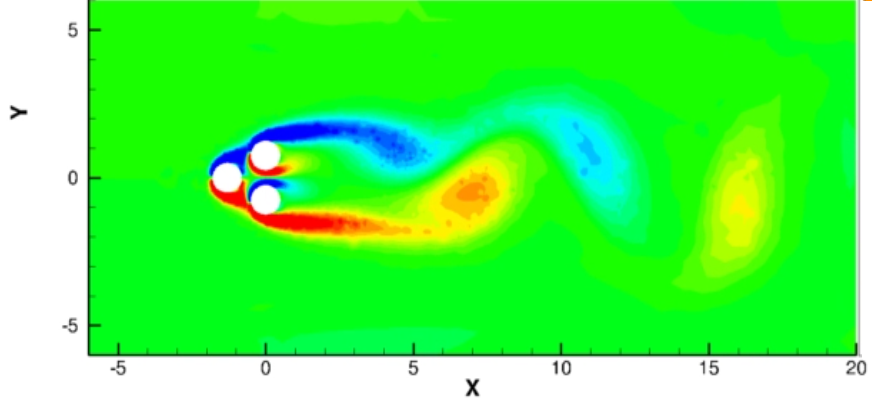}
 \\
 (b) \\ \includegraphics[width=\linewidth]{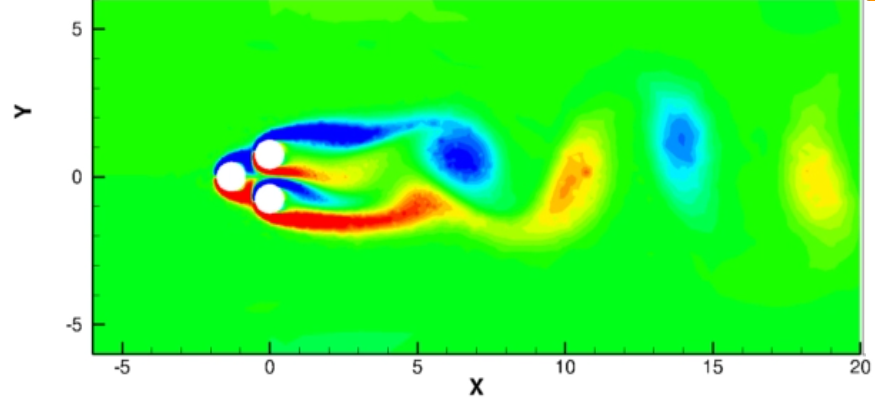}
 \\
\end{tabular}
 \caption{INSTANTANEOUS SNAPSHOTS OF THE NATURAL FLOW AT $Re_D=50$ (a), $Re_D=90$ (b). THE COLORMAP ENCODES THE VORTICITY FIELD.} 
 \label{fig:jet}
\end{figure}

\begin{figure}[tb]
\begin{center}
 \includegraphics[width=00.45\textwidth]{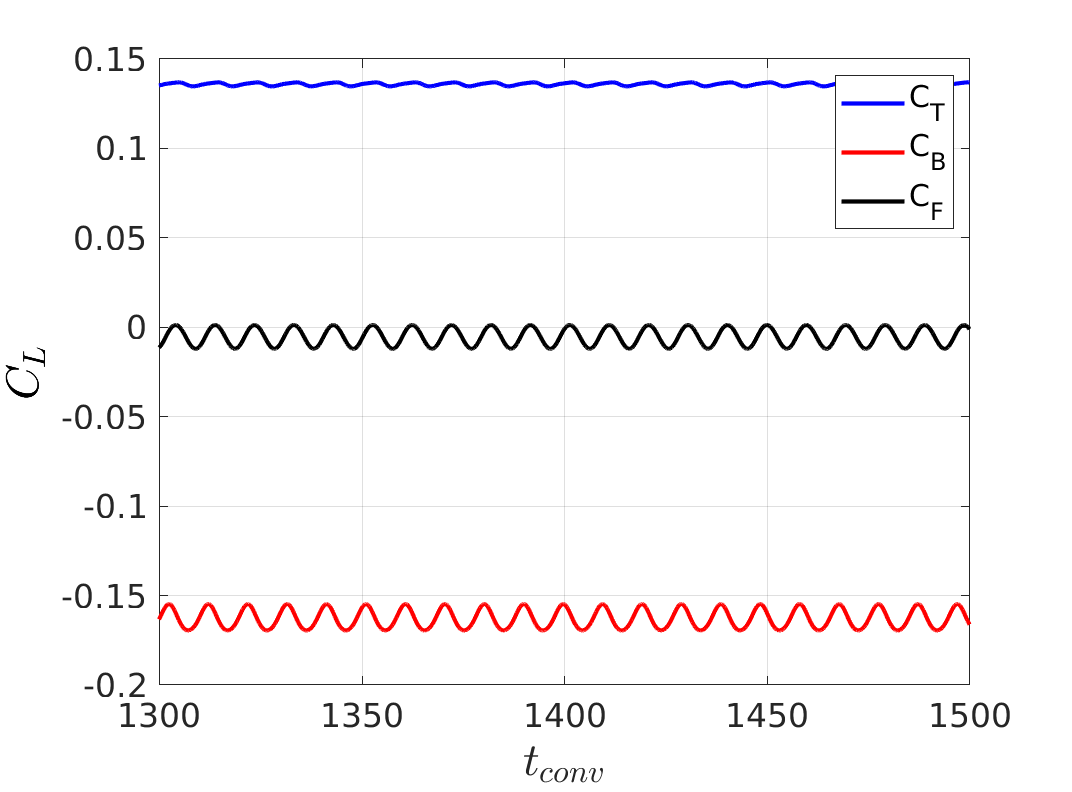}   
\caption{TIME SERIES OF THE LIFT COEFFICIENTS ON DIFFERENT CYLINDERS  AT $Re_D=75$.}
\label{fig:ts75} 
\end{center}
\end{figure}

\begin{figure}[tb]
\begin{center}
 \includegraphics[width=0.45\textwidth]{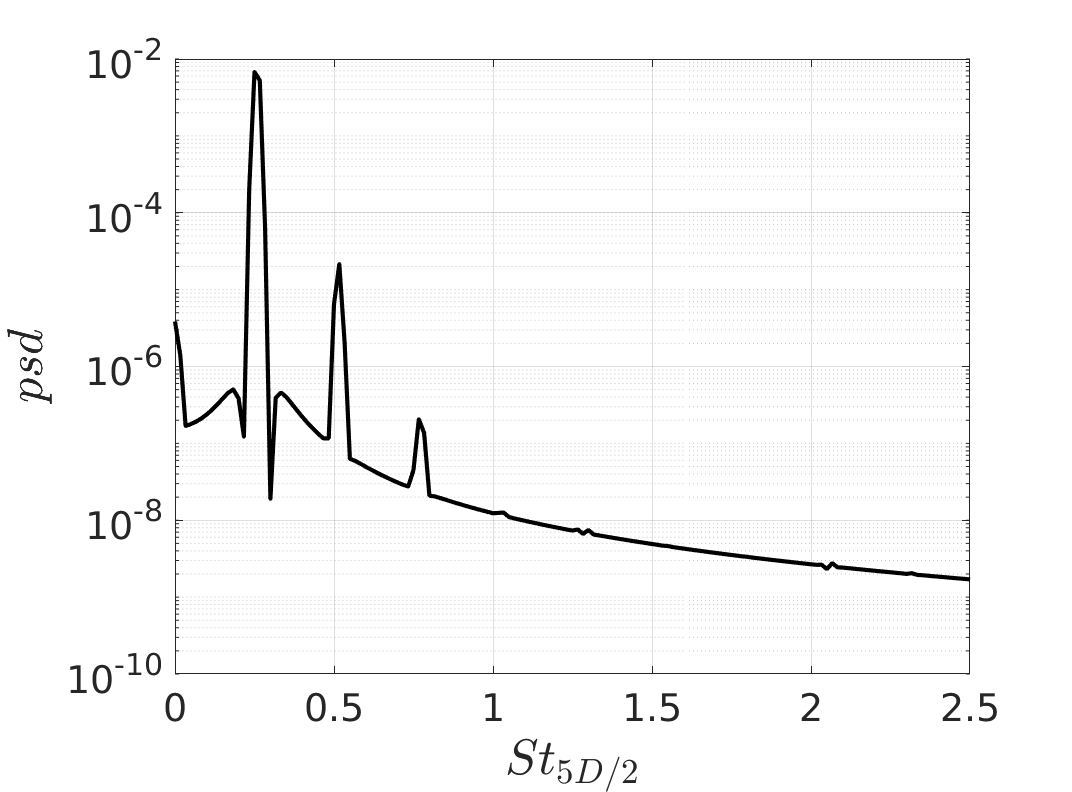}   
\caption{POWER SPECTRAL DENSITY OF THE LIFT COEFFICIENT AT $Re_D=75$. STROUHAL NUMBER BASED ON $5D/2$ FOR COMPARISON WITH THE SINGLE CYLINDER WAKE FLOW.}
\label{fig:psd75} 
\end{center}
\end{figure}

\begin{figure}[tb]
\begin{center}
 \includegraphics[width=00.45\textwidth]{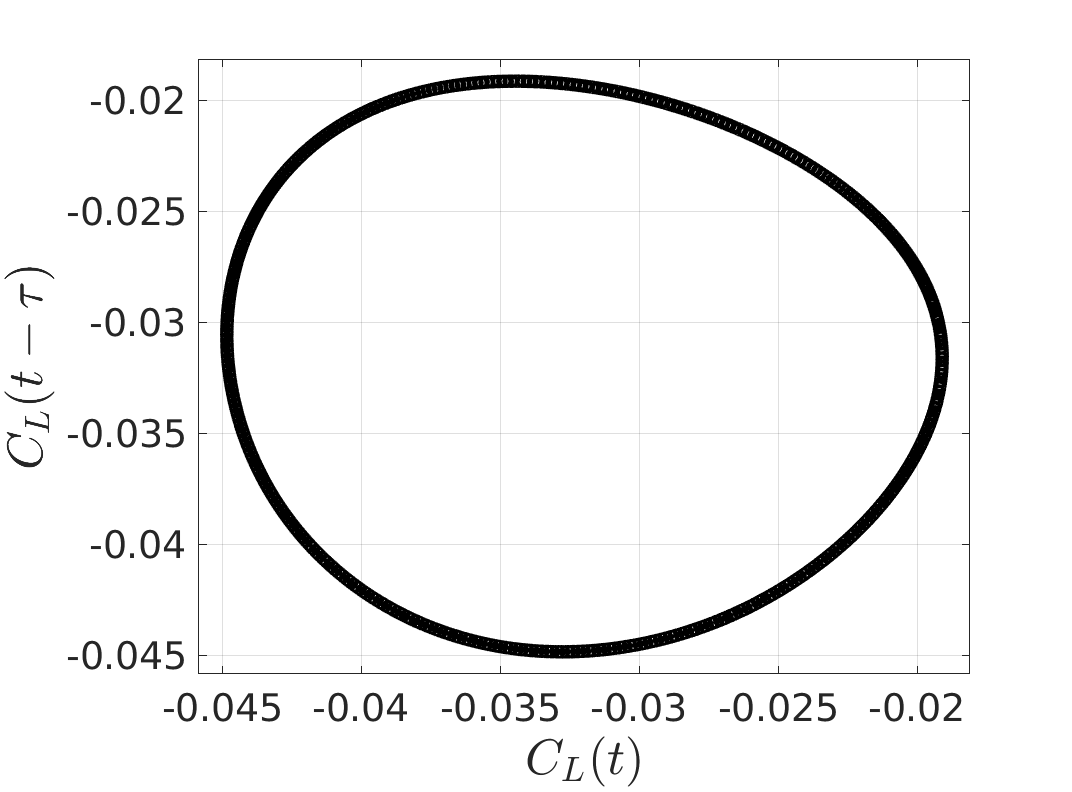}  
\caption{PHASE PORTRAIT BASED ON THE LIFT COEFFICIENT AT $Re_D=75$. TIME DELAY $\tau =1/(4f)$ WHERE $f$ IS THE FREQUENCY OF THE DOMINANT PEAK IN THE POWER SPECTRUM.}
\label{fig:portrait75} 
\end{center}
\end{figure}

In this section, we shall focus on the route to chaos undergone by the fluidic pinball in the natural regime, i.e. without actuation, as the Reynolds number is smoothly increased. To that aim, let consider the lift coefficient $C_L=2F_L/\rho U_\infty ^2$, where $F_L$ is the lift force applying to the cylinders in direction of $y$-axis. In the forthcoming part of the paper, we shall also be interested in individual lift coefficients acting on each individual cylinders, namely $C_{F}$, $C_{T}$, $C_{B}$ when considering the front, top or bottom cylinders, respectively.

The steady solution, shown in figure~\ref{fig:steadysolution} for $Re_D=10$, is stable up to the critical value $Re_{c1}\simeq 18$ of the Reynolds number (the critical value would be about 45 in units of $L$). Beyond this value, the lift coefficient becomes oscillatory, as indicated by its non vanishing standard deviation shown in Fig.~\ref{fig:CL} (dashed line). The system has undergone a supercritical Hopf bifurcation characterized in the flow field by the usual vortex shedding phenomenon and generation of the von K\'arm\'an vortex street, shown in Fig.~\ref{fig:jet}(a) for $Re_D=50$. The jet formed at the base of the two outer cylinders, as shown in Fig.~\ref{fig:jet}(a), absent from the single cylinder wake, has a critical impact on the successive bifurcations undergone by the system on the route to chaos. Indeed, at a secondary critical value of the Reynolds number $Re_{c2}\simeq 68$, the system undergoes a new bifurcation that breaks the symmetry of the mean flow field. In figure~\ref{fig:jet}(b), the base-bleeding jet appears to be deflected to the bottom, a feature absent from figure~\ref{fig:jet}(a), when $Re<Re_{c2}$. As a result, the mean value of the lift coefficient is non-vanishing anymore, as shown in Fig.~\ref{fig:CL} (solid line). This symmetry-breaking in the mean flow field is typical of a pitchfork bifurcation, on the top of which develops the oscillatory behavior inherited from the primary Hopf bifurcation. This is also exemplified in Fig.~\ref{fig:ts75} where both the top $C_T$ and bottom $C_B$ lift coefficients are plotted as functions of the convective time $t_{\mathrm{conv}}$ (expressed in units of $D/U_\infty$). Both lift coefficients oscillate at the same frequency but with two different amplitudes, the lift at the bottom cylinder being much stronger than the lift at the top cylinder. 
The main peak in the power spectrum of the lift coefficient is found at $St_D = \frac{fD}{U} \simeq 0.10$, together with its first harmonic, corresponding to a Strouhal number of about 0.25 in units of $5D/2$ (units used in the power spectrum of Fig.~\ref{fig:psd75}). The phase portrait built on the lift coefficient exhibits a limit cycle, as expected after a primary Hopf bifurcation, see Fig.~\ref{fig:portrait75}. At the precision of our investigation, both the Hopf and pitchfork bifurcations were found to be supercritical. 

\begin{figure}[tb]
\begin{center}
 \includegraphics[width=00.45\textwidth]{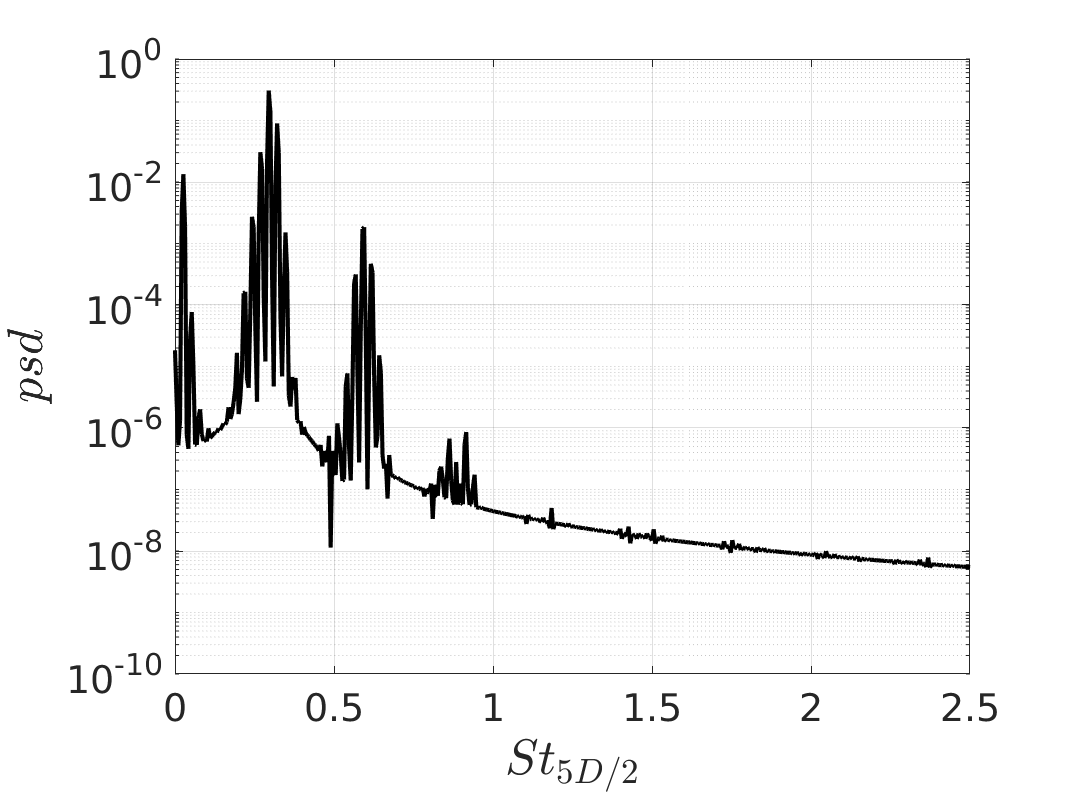}   
\caption{POWER SPECTRAL DENSITY OF THE LIFT COEFFICIENTS AT $Re_D=105$. STROUHAL NUMBER BASED ON $5D/2$ FOR COMPARISON WITH THE SINGLE CYLINDER WAKE FLOW.}
\label{fig:psd105} 
\end{center}
\end{figure}

\begin{figure}[tb]
\begin{center}
 \includegraphics[width=00.45\textwidth]{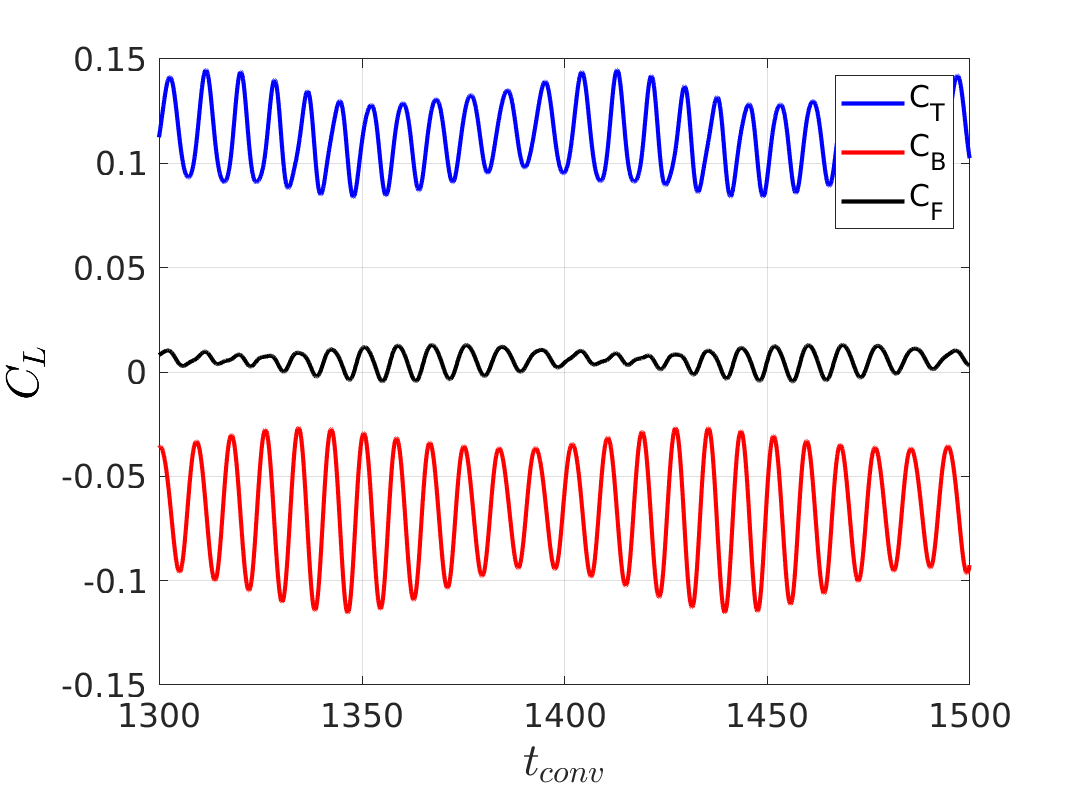}   
\caption{TIME SERIES OF THE LIFT COEFFICIENTS ON DIFFERENT CYLINDERS AT $Re_D=105$.}
\label{fig:ts105} 
\end{center}
\end{figure}

\begin{figure}[tb]
\begin{center}
 \includegraphics[width=00.45\textwidth]{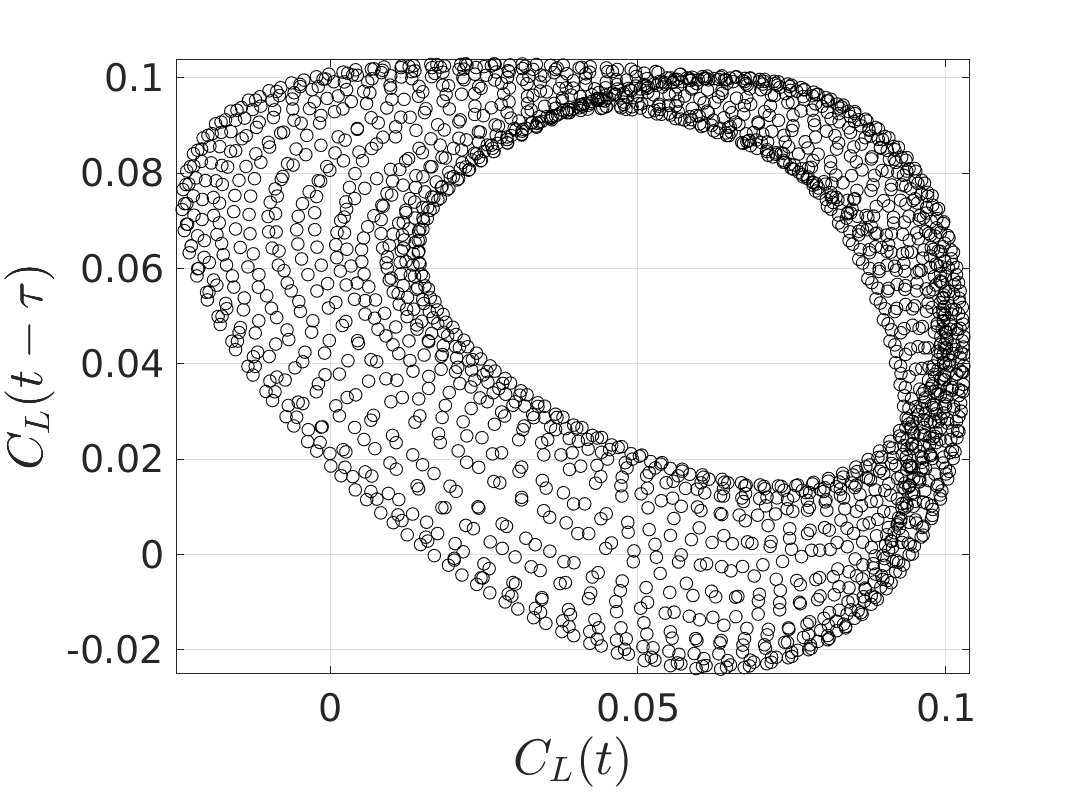}  
\caption{PHASE PORTRAIT BASED ON THE LIFT COEFFICIENT AT $Re_D=105$. TIME DELAY $\tau =1/(4f)$ WHERE $f$ IS THE FREQUENCY OF THE DOMINANT PEAK IN THE POWER SPECTRUM.}
\label{fig:portrait105} 
\end{center}
\end{figure}

When the Reynolds number is further increased up to a critical value $Re_{c3}\simeq 104$, a new frequency ($St_D \simeq 0.0105$) rises in the power spectrum. This frequency is about one order of magnitude smaller than the natural frequency of the vortex shedding ($St_D \simeq 0.118$), as illustrated in figure~\ref{fig:psd105} for $Re=105$ (spectral component closest to the vertical axis). A visual inspection of both the time series (Fig.~\ref{fig:ts105}) and the phase portrait (Fig.~\ref{fig:portrait105}), at $Re=105$, indicates that the new frequency also modulates the amplitude of the main oscillator and thickens the limit cycle associated with the main oscillator (figure~\ref{fig:portrait105}). All those features are typical of a quasi-periodic dynamics, indicating that the system has most likely undergone a Neimark-S\"acker bifurcation at $Re\simeq Re_{c3}$ \cite{guckenheimer2013}.

\begin{figure}[tb]
\begin{center}
 \includegraphics[width=00.45\textwidth]{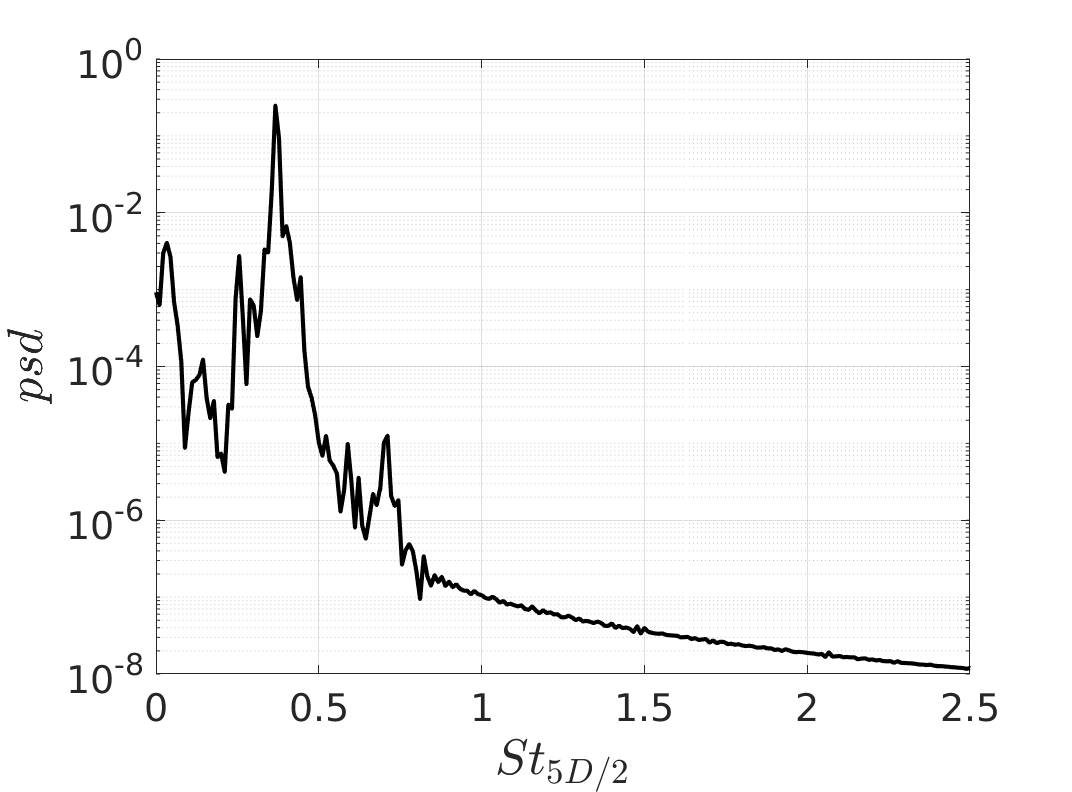}   
\caption{POWER SPECTRAL DENSITY OF THE LIFT COEFFICIENTS AT $Re_D=120$. STROUHAL NUMBER BASED ON $5D/2$ FOR COMPARISON WITH THE SINGLE CYLINDER WAKE FLOW.}
\label{fig:psd120} 
\end{center}
\end{figure}

\begin{figure}[tb]
\begin{center}
 \includegraphics[width=00.45\textwidth]{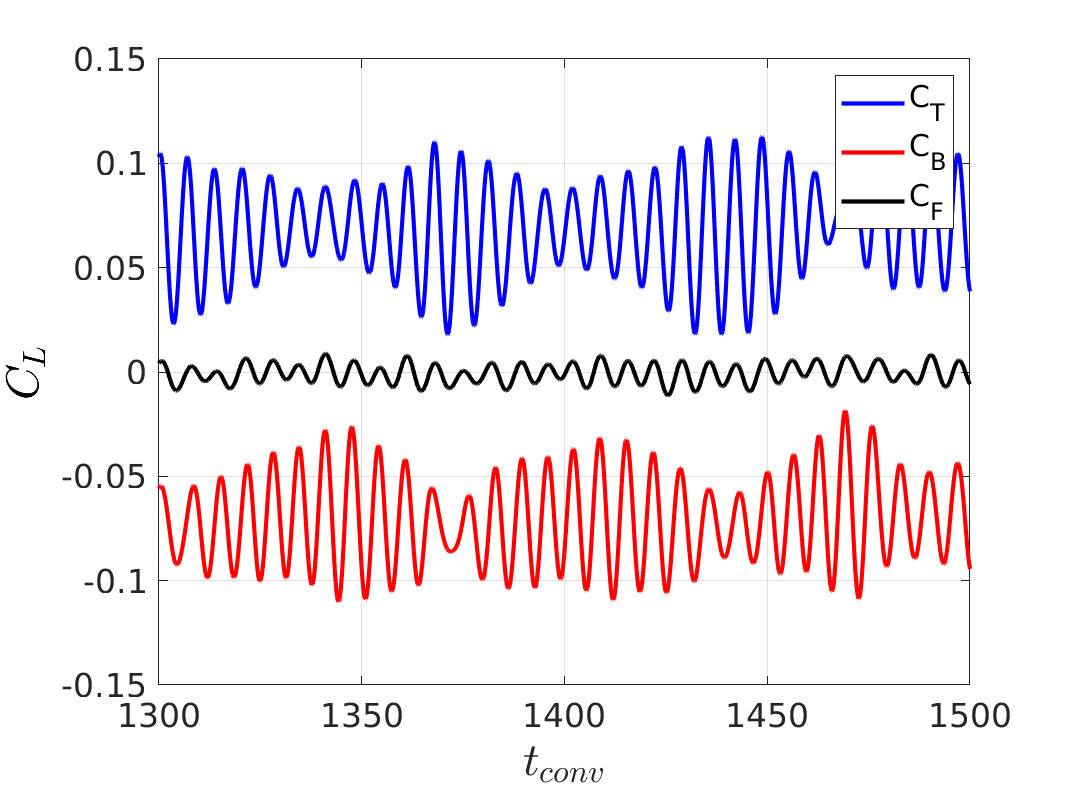}   
\caption{TIME SERIES OF THE LIFT COEFFICIENTS ON DIFFERENT CYLINDERS AT $Re_D=120$.}
\label{fig:ts120} 
\end{center}
\end{figure}

\begin{figure}[tb]
\begin{center}
 \includegraphics[width=00.45\textwidth]{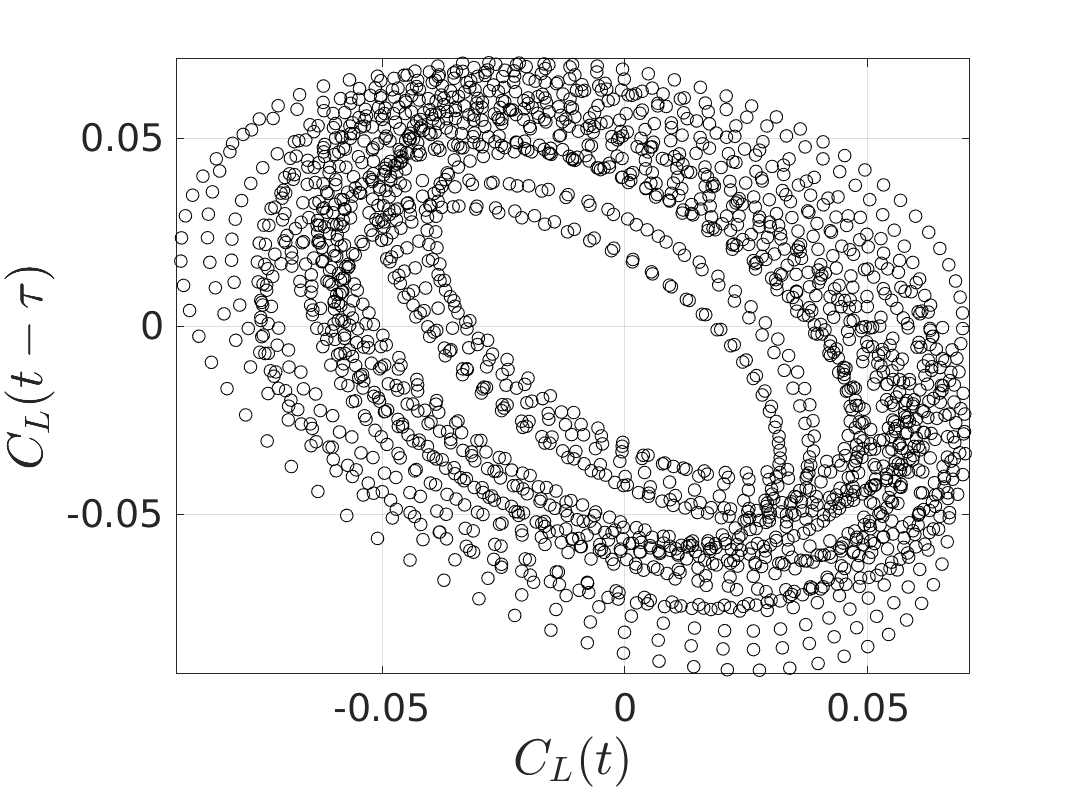}  
\caption{PHASE PORTRAIT BASED ON THE LIFT COEFFICIENT AT $Re_D=120$. TIME DELAY $\tau =1/(4f)$ WHERE $f$ IS THE FREQUENCY OF THE DOMINANT PEAK IN THE POWER SPECTRUM.}
\label{fig:portrait120} 
\end{center}
\end{figure}

At even larger values of the Reynolds number, the main peak in the power spectral density of the lift coefficient widens significantly, as shown in figure~\ref{fig:psd120} for $Re_D=120$. The time series do not exhibit neither periodic nor quasi-periodic features anymore (see figure~\ref{fig:ts120}) and the phase portrait exhibits a much more complex dynamics (see figure~\ref{fig:portrait120}). The dynamical regime, henceforth, exhibits many features of a chaotic regime, indicating that the system has followed the Ruelle-Takens-Newhouse route to chaos \cite{newhouse1978}.

\section*{DISCUSSION}

The symmetry-breaking observed in the wake flow of the fluidic pinball, on the route to chaos, is something that is usually not observed in the wake of a single cylinder flow. Instead, it looks more similar to what was recently discovered in the wake of three-dimensional bluff bodies, see for instance \cite{grandemange2013a,grandemange2013b,rigas2014,cadot2015,barros2016}. In squared-base bluff bodies, bi-stable dynamics are observed in which the wake flow randomly switches from one deflected position to the conjugated one \cite{grandemange2013a,grandemange2013b,barros2016}.  It can be expected that the dynamics in the fluidic pinball be also bi-stable beyond the pitchfork bifurcation that breaks the symmetry of the mean flow, since two mirror conjugated wake flows must exist, which are perturbed by the unsteadiness of the von Karman street. This question remains open to further investigations. 

Beyond its apparent simplicity, the \textit{fluidic pinball} provides a most convenient sandbox to investigate fundamental aspects of fluid flow dynamics: 
\begin{itemize}
 \item \textbf{Model reduction}. The seeding paper by Noack \textit{et al} 2003 \cite{noack2003} proved that the von K\'arm\'an vortex street dynamics past a cylinder could be reduced to a non-linear dynamical system of at least three degrees of freedom, in the frame of POD-Galerkin projections of the Navier-Stokes equations. Loiseau \textit{et al} 2017 \cite{loiseau2017} provided a 4 dimensional reduced-order model for the fluidic pinball at $Re_D=60$, running a SINDy identification \cite{brunton2016} on the dynamics of each individual lift coefficients $C_F$, $C_T$, $C_B$. One challenging task would be to provide such a reduced-order model beyond the secondary pitchfork bifurcation, where additional degrees of freedom are expected to occur. Another step would be to generalize the model over a wide range of the Reynolds number by determining the Reynolds dependance of the model coefficients, say from 0 to $Re_{c3}$, or even beyond. These are works under progress. \\
 
 \item \textbf{Machine learning control.}  As already stressed out in the introductory part, designing MLC on a numerical flow configuration must rely on fast enough computational flow dynamics such as to allow testing hundreds or thousands of individual control laws \cite{duriez2017}. The computational time for vortex shedding in the \textit{fluidic pinball} is comparable to the characteristic time of vortex shedding in its experimental counterpart, which makes it competitive with respect to an experimental setup. 
\begin{figure}[h]
\begin{tabular}{c}
 \includegraphics[width=\linewidth]{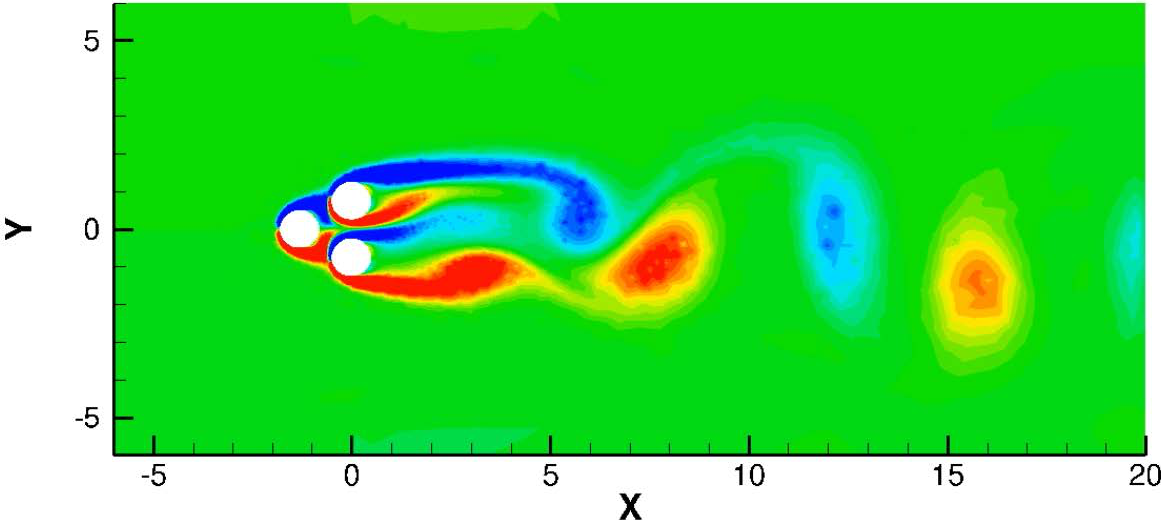} \\
 \includegraphics[width=\linewidth]{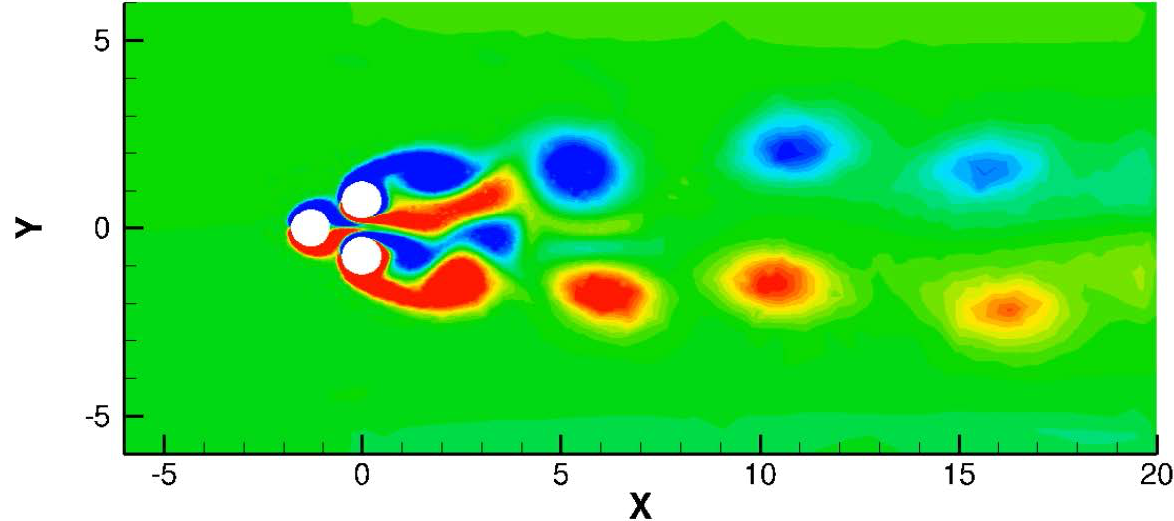} \\
 \includegraphics[width=\linewidth]{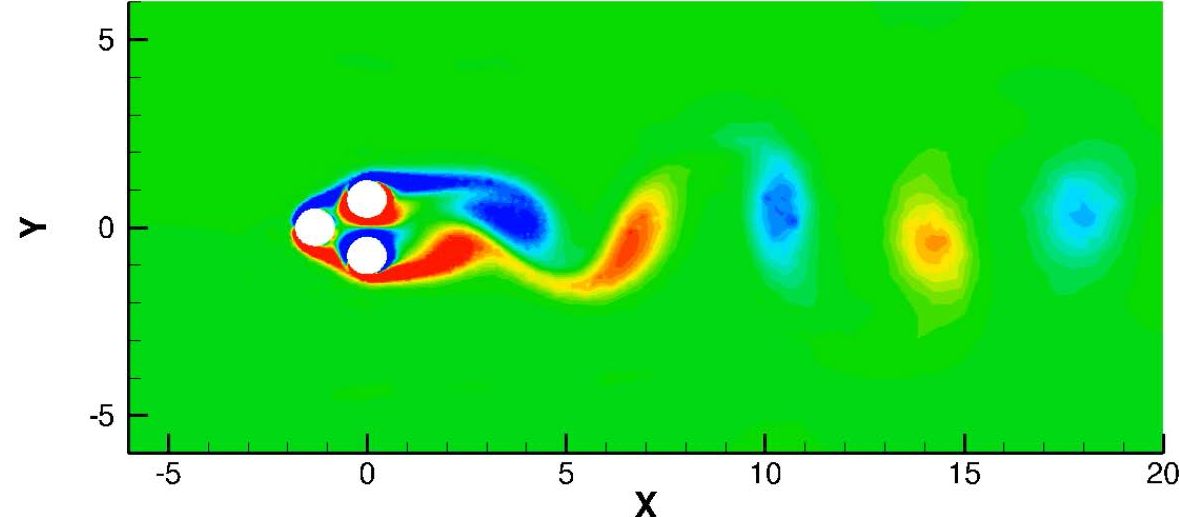} \\
 \end{tabular}
 \caption{NATURAL FLOW (a), BASE BLEED (b) AND BOAT TAILING ACTUATIONS (c), IN THE PERMANENT REGIME AT $Re_D=105$ (SNAPSHOTS AT AN INSTANTANEOUS TIME). MAGNITUDE OF THE ROTATION SPEED OF THE RIGHTMOST CYLINDERS IS $2U_\infty$. THE COLORBAR IS IDENTICAL IN EACH SNAPSHOT. }
 \label{fig:actuation}
\end{figure}
As reported in \cite{tutorial}, base bleed and boat tailing both provide efficient mechanisms to act on the flow, among other strategies. Base bleed consists in rotating at a constant speed the two rightmost cylinders symmetrically with respect to the horizontal axis, the top cylinder rotating in the anti-clockwise direction and the bottom cylinder in the clockwise direction. The physical effect of base bleed is similar to the splitter plate. The communication between the upper and lower shear layer is suppressed. Thus, a von K\'arm\'an vortex cannot occupy the whole near wake region but is pushed away in its infancy \cite{wood1964,bearman1967}. Boat tailing is the base bleed symmetrical action, where the top cylinder turns in the clockwise direction at a constant speed (anti-clockwise for the bottom cylinder). With boat tailing, the drag can also be reduced by shaping the wake region more aerodynamically, \textit{i.e.} by vectoring the shear layer towards the center region. This has been performed by passive devices, like vanes \cite{flugel1930}, or active control via Coanda blowing \cite{geropp2000,barros2016}. Examples of base bleed and boat tailing on the fluidic pinball are shown in Fig.~\ref{fig:actuation} (b) and (c), respectively, together with the natural flow (a), at $Re_D=105$.

 \item 
\end{itemize}

\section*{CONCLUSION}

The \textit{fluidic pinball} provides an effective sandbox for reduced-order modeling of fluid flows and design of real-time fluid flow control. As reported in this paper, the route to chaos in the natural flow regime, without actuation, when the Reynolds number is increased, is of the Newhouse-Ruelle-Takens kind \cite{newhouse1978}, with a secondary pitchfork bifurcation that breaks the symmetry of the mean flow field on the route to quasi-periodicity. Forthcoming studies will rely on both challenging aspects of reduced-order modeling and machine learning flow control in tractable numerical flow simulations.

\bibliographystyle{asmems4}

\begin{acknowledgment}
This work is part of a larger project involving S. Brunton, G.~Cornejo Maceda, J.C. Loiseau, F. Lusseyran, R.~Martinuzzi, C.~Raibaudo, R.~Ishar and many others. 

This work is supported by the ANR-ASTRID project ``FlowCon'',  by a public grant overseen by the French National Research Agency (ANR) as part of the ``Investissement d'Avenir'' program, ANR-11-IDEX-0003-02, and by the Polish National Science Center (NCN) under the Grant No.: DEC-2011/01/B/ST8/07264 and  by the Polish National Center for Research and Development under the Grant No. PBS3/B9/34/2015.

\end{acknowledgment}

%
% ...

\bibliography{asme2e}


\end{document}